\documentclass[aps,amsfonts,amsmath,prd,preprint,nofootinbib]{revtex4}
\usepackage{epsf}

\newcommand{\beq}{\begin{equation}}
\newcommand{\eeq}{\end{equation}}

\begin{document}

\title{Prediction and explanation in the multiverse}

\author{J. Garriga$^1$ and A. Vilenkin $^2$}
\address{
$^1$ Departament de F{\'\i}sica Fonamental, Universitat de Barcelona,\\
Mart{\'\i}\ i Franqu{\`e}s 1, 08193 Barcelona, Spain\\
$^2$ Institute of Cosmology, Department of Physics and Astronomy,\\
Tufts University, Medford, MA 02155, USA}

\begin{abstract}

Probabilities in the multiverse can be calculated by assuming that we
are typical representatives in a given reference class. But is this
class well defined? What should be included in the ensemble in which
we are supposed to be typical? There is a widespread belief that this
question is inherently vague, and that there are various possible
choices for the types of reference objects which should be counted
in. Here we argue that the ``ideal'' reference class (for the purpose
of making predictions) can be defined unambiguously in a rather
precise way, as the set of all observers with identical information
content. When the observers in a given class perform an experiment,
the class branches into subclasses who learn different information
from the outcome of that experiment. The probabilities for the
different outcomes are defined as the relative numbers of observers in
each subclass. For practical purposes, wider reference classes can be
used, where we trace over all information which is uncorrelated to the
outcome of the experiment, or whose correlation with it is beyond our
current understanding.  We argue that, once we have gathered all
practically available evidence, the optimal strategy for making
predictions is to consider ourselves typical in any reference class we
belong to, unless we have evidence to the contrary.  In the latter
case, the class must be correspondingly narrowed.

\end{abstract}

\maketitle

\section{Introduction}

The task of assigning probabilities to the outcomes of observations
has a long tradition in physics, and we are used to the fact
that such predictions improve as we progress in our understanding of
the fundamental theory.  Yet, the current situation in cosmology is
paradoxical.  The inflationary scenario, which is now the leading
cosmological paradigm, suggests a very large universe with a variety
of different environments.  For example, the density fluctuations in
this scenario are stochastic variables determined by quantum
fluctuations of the inflaton field.  The fluctuation amplitudes and
spectra are different in different parts of the universe.
In axion models of dark matter, the dark matter density is also a
random variable, determined by the local amplitude of the axion field.
In this situation, it seems pertinent to ask the following question.
Given some information about the properties of our own environment,
can we assign probabilities to any of the remaining properties?
Despite its simplicity, there is no consensus on how this question
should be tackled, or whether it has an answer at all.

The problem is particularly acute when the fundamental theory admits a
multitude of solutions describing vacua with different values of the
low-energy constants of nature, as may be the case in string theory
\cite{landscape}. In the cosmological context, high-energy vacua drive
an exponential inflationary expansion of the universe.  Transitions
between different vacua can occur through tunneling and quantum
diffusion, so ``pocket universes'' filled with all possible kinds of
vacua are constantly being formed and the entire landscape of vacua
can be explored \cite{inflation}.  In this scenario, the local values
of the constants cannot be predicted with certainty, but we may be
able to make statistical predictions.  The prediction of the observed
value of the cosmological constant
\cite{Weinberg89,AV95,Efstathiou,MSW} is a notable success of this
approach, but the methods used for that prediction have been a subject
of much recent criticism \cite{HS,Starkman,Krauss}.

In anthropic predictions of this sort one usually assumes that we are
typical representatives in some reference class of observers, and the
arbitrariness in the choice of the reference class is the main target
of the criticism.  Should the reference class include all living
intelligent creatures?  If so, how do we define life and intelligence?
Should chimpanzees be included?  And whatever our choice of reference
class, how do we know that we are typical in that class?

This situation is to be contrasted with that in quantum mechanics,
where we imagine an infinite ensemble of identical experiments
performed by identical observers.  (The observers are actually existing
in the many worlds interpretation.)  Once the experiment is performed,
the ensemble branches into a number of sub-ensembles corresponding to
the different outcomes of the experiment, and the probabilities are
identified with the corresponding branching ratios.

This paper is an attempt to clarify the definitions of probability and
reference class in multiverse models.  We shall argue that, despite a
widespread belief to the contrary, these concepts can be defined
unambiguously in a rather precise way (which is somewhat analogous to
the one based on the branching ratios in quantum mechanics).

We should mention that there is a further complication in the
case of the inflationary multiverse. The problem is that the number of
elements in any given reference class is infinite, and the
relative number of elements in each subclass depends on how we
regulate the infinities. This is known as the measure problem, which is
currently a subject of active research (for recent discussions,
see \cite{GSPVW,Linde07,Tegmark,Bousso,cfh,AVreview}). Here we shall not
discuss this problem any further, since we are addressing a different set of issues
which are relevant to the results obtained from any given regulator.

\section{The branching reference class}

For any given event, eternal inflation leads to an infinite number of
occurrences, coexisting in different regions of a single spacetime
continuum.  In the inflationary "multiverse" there are an infinite
number of pocket universes of any given type.  And if an event can
occur inside of a given pocket, then this pocket will contain not just
one but an infinite number of such events.

Suppose a measurement ${\cal M}$ is about to be made, which can yield
a number of possible datasets $D_j$, and some ``observer'' (not
necessarily a member of the observational team) wants to predict the
outcome of this measurement.  We can characterize the observer by the
information she has that is relevant for the experiment, as well as by
all the irrelevant information, such as her address, the name of her
dog, etc.  In other words, the observer is characterized by the full
information content at her disposal.  The key point is that there is
an infinite number of observers with an identical information content
in the multiverse \cite{GV00}.  We suggest that, ideally,
this set of observers should be chosen as the reference class $C$, to
be used for making predictions.  (A much wider class is usually
used for practical considerations; see below.)

Once the measurement is made, observers get new information from the
data, and the class $C$ splits into a number of subclasses $C_j$
corresponding to different outcomes $D_j$.  The observer information
content is the same within each subclass, but differs from one
subclass to another.  The probability of a dataset $D_j$ should be
identified with the fraction $f_j$ of observers who end up in
subclass $C_j$, 
\beq
P(D_j)=f_j.
\label{Pf}
\eeq 
This equation assigns equal statistical weight to all observers
in class $C$.  In other words, we think of ourselves as being randomly
picked from that class of observers.\footnote{The proposal of using
the full information content of observers for making predictions has
also been advanced by Neal \cite{Neal}.  However, he assumed a finite
universe, small enough, so that observers can be uniquely specified by
their information content.  This is very different from the situation
we are discussing here.}  This is justified by the fact that all
observers in $C$ have identical information content and thus have no
way of distinguishing between one another.\footnote{The actual
calculation of the probabilities is complicated by the fact that the
number $N$ of observers in $C$ is infinite.  In order to calculate the
fractions $f_j$, one has to impose some sort of a cutoff on $N$ and
then remove the cutoff at the end of the calculation.  The result,
however, turns out to be sensitive to the cutoff procedure. This is
the so-called measure problem which we refered to at the end of the
Introduction.  Since our considerations apply to the results obtained from any
cut-off procedure, we shall not dwell on this problem any further.}

If another measurement ${\cal M'}$ is made after ${\cal M}$, the
probabilities for its possible outcomes ${D'}_j$ should be updated
using the Bayes formula,
\beq
P({D_j}'|D_i)={P({D_j}',D_i)\over{P(D_i)}}.
\label{bayes}
\eeq
The probabilities for ${D'}_j$ assigned by observers in different
subclasses $C_j$ are, of course, generally different.  Every
subsequent measurement will lead to a further splitting of reference
classes.  The resulting ``branching tree'' of reference classes is
similar to the branching wave function in the many worlds
interpretation of quantum mechanics.

It should be stressed that the above discussion assumes that we
have a theory allowing us to calculate the probabilities $P(D_j)$.
Agreement (or lack thereof) of these probabilities with the data can
be used to evaluate the theory, as we shall discuss below, in Section \ref{evidence}

\section{The principle of mediocrity}

Using the full information content of observers for calculation of
probabilities is not a very practical proposition.  One only needs to
use the relevant information which has non-negligible correlation with
the data.  Moreover, our models of the multiverse are too crude and
our understanding of life and intelligence too rudimentary to account
for more than very basic information.  Thus, by necessity, we need to
consider a reference class of observers specified by a small subset of
all available information, and thus much wider than the class $C$
defined above.

Suppose, for example, we want to predict the sum of the three neutrino
masses, $m_\nu$.  (The mass differences are already known from
neutrino oscillation measurements.)  For this purpose, we could
consider the reference class $C^{(\nu)}$ of all human-like observers
who measured the same values for all constants of nature other than
$m_\nu$.  The probability distribution $P(m_\nu)$ can then be
calculated \cite{Levon}, with some additional assumptions about the
landscape of possible vacua.  This distribution gives the probability
that an observer randomly picked in the class $C^{(\nu)}$ will
measure a given value of $m_\nu$.  Serious doubts have been raised,
however, that such distributions can be used for making predictions in
our local region \cite{HS,Starkman,Krauss}.

The class $C^{(\nu)}$ is specified by very limited information, which
does not include the amplitudes of the CMB multipoles, observer's
addresses, etc.  Much of the omitted information is irrelevant, that
is, has negligible correlation with the value of $m_\nu$.  But some
information may be relevant (e.g., the abundance of galaxy clusters is
known to be correlated with $m_\nu$), and for some other information
the relevance may be hard to assess.  Why, then, should we think of
ourselves as typical (randomly selected) in the reference class
$C^{(\nu)}$?  This is what Hartle and Srednicki \cite{HS} call
``the selection fallacy''.

Hartle and Srednicki have argued that we should never assume ourselves
to be typical representatives of some reference class of observers,
unless we have evidence to back up that assumption.  We
disagree.\footnote{See also the article by Page \cite{Page2} for a
critique of Hartle and Srednicki.}  We would suggest that, on the
contrary, we should assume ourselves to be typical in any class that
we belong to, unless there is some evidence to the contrary.  This is
a statement of the principle of mediocrity.\footnote{The original
formulation of the principle of mediocrity in \cite{AV95} asserts that
we should regard our civilization as randomly picked among all
civilizations in the multiverse.  Similar ideas have been earlier
discussed by Gott \cite{Gott} and Leslie \cite{Leslie} and about the
same time by Page \cite{Page}.  Our formulation here is more general.
A related proposal is the ``self-sampling assumption'' introduced by
Bostrom \cite{Bostrom}: ``One should reason as if one were a random
sample from the set of all observers in one's reference class.''  The
problem with this formulation though is that it does not specify how
the reference class is to be selected.  We should also mention
Carter's anthropic principle \cite{Carter}, which served as an
inspiration for all these ideas.}

Even though observers in class $C^{(\nu)}$ may suspect that some
relevant information may be missing, the principle of mediocrity
recommends that they should make their predictions using the
probability distribution for that class.  When the missing information
becomes available, the class $C^{(\nu)}$ will split into subclasses
${C^{(\nu)}}'$, ${C^{(\nu)}}''$, etc.  Observers in $C^{(\nu)}$ would
be able to make more accurate predictions if they could calculate the
distributions for their respective subclasses.  But in the absence of
such distributions, they should base their bets on the distribution
for $C^{(\nu)}$.  Some of them will guess incorrectly, but there will
be more winners than losers.  This is the justification for the
principle of mediocrity: it simply improves the odds.

The principle of mediocrity is widely used in physics, although it is
rarely acknowledged explicitely.  For example, an experimentalist
representing the result of a measurement as $x\pm \delta x$ thinks of
his data as a random sample drawn from a Gaussian distribution of
width $\delta x$.  Similarly, cosmologists think of the CMB multipoles
as randomly drawn from Gaussian distributions.  Without assuming that
the data are typical in this sense, we would have to conclude that the
measurement gives us no information about the quantity being measured.

\section{Indexical information}

An experiment that is very easy for anyone to do is self-inspection,
by which one can recognize that one exists and belongs to a certain
group.  The result of this experiment is usually called indexical
information, and may sometimes influence the probabilities
(\ref{bayes}) \cite{Bostrom2}.

To illustrate the last point, consider a special case when all
observers share the same information, except for the indexical one.
Suppose we have a universe populated by civilizations that can be
either large, with a number of citizens $N=L$ or small, with $N=S$.
These civilizations are peculiar in the following respects. First, by
assumption, nobody in any of the civilizations has been able to count,
or even estimate, how many individuals belong to their own
civilization. Let us assume that for some reason this was hard to
do. However, it suddenly becomes known that a given type of
cosmological observation, which we call ${\cal O}$, is completely
correlated with the number of people in the civilizations, so that
$D_{\cal O}=N$.  Second, it is also known that there is just one
astronomer in each civilization, who performs the observation ${\cal
O}$. After obtaining the result, the astronomer makes it available to
the rest of citizens in that particular civilization. Suppose everyone
involved is perfectly aware of all the facts just listed, plus the
additional fact that the number of small and large civilizations
in the universe is the same.
Given this information, the astronomers performing the
observations should rightly bet that there is a 50 \% chance that the
observation will yield $D_{\cal O}=L$ and a 50\% chance that it will
yield $D_{\cal O}=S$.  It may therefore seem surprising that the plain
citizens, who learn about the outcomes of observations by the very
astronomers we just discussed, should place their bets differently.

Indeed, a plain citizen can reason as if he has been randomly picked
from all the people in the multiverse, and thus he is more likely to
be in a large civilization than in a small one.  Hence, the plain
folks should assign probabilities
\begin{equation}
P(D_{\cal O}=L)={L\over L+S},\quad P(D_{\cal O}=S)={S\over
L+S},\label{pcit}
\end{equation}
This may seem paradoxical, since astronomers and plain folks seem to
share the same information.  However, this assessment is deceptive.
The astronomer knows that she exists as an astronomer, whereas the
citizen knows that he exists as such. Because of this difference in
the indexical information 
their expectations for the outcome of the same experiment are, and
should be, different. 

Even if the astronomer, wiser than the citizens, would have to give
them advice, this would have to be to use the probabilities
(\ref{pcit}). This may sound hypocritical, when the astronomer herself
is using 1/2 and 1/2. However, after the experiment, the civilizations
come to know their numbers. In small civilizations, the citizens would
have done worse than the astronomer, whereas in large civilizations
they would have turned out to do better.  Hence, the astronomer
minimizes liabilities by advising the citizens to predict according to
(\ref{pcit}).\footnote{We note that indexical information can be
transfered, just like all other forms of information.  If, for
instance, the astronomer has a "friend" whom she singles out of the
crowd and with whom she communicates, then indexical information is
transfered from the astronomer to that particular individual.  This
individual will reason that he is unlikely to have been picked by the
astronomer from a large group, and will update his probabilities
accordingly. As a result, his expectation for the outcome of the
experiment will coincide with that of the astronomer. In our idealized
example we are assuming that there is no exchange of indexical
information prior to the experiment.}  

How should we then assign probabilities?  Should we identify with
astronomers or with regular folks?  Both identifications have
analogues in the definitions of probabilities that have been discussed
in the literature.  The probability assigned to a given environment
has been defined as the probability for a randomly picked (i)
civilization \cite{AV95}, (ii) observer \cite{Gott,Leslie}, (iii)
observation \cite{Page} or (iv) ``observer-moment'' \cite{Bostrom} to
be in that environment.  The meaning of ``observation'' in \cite{Page}
and ``observer-moment'' in \cite{Bostrom} is essentially the same as
the information content of an observer, so there is not much
difference between (ii), (iii) and (iv).  But the choice between
civilizations and observers is similar to that between astronomers and
regular folks in our example.

The motivation for using civilizations is that modern astronomical and
particle physics measurements are collective enterprises, requiring
the resources of the entire civilization.  Moreover, once the
measurement is made, one can assume that the result becomes available
to all interested citizens.  As our example shows, predictions based
on the reference class of civilizations may differ from those using
the reference class of individuals.  In such cases, the same
person may wish to bet differently as an individual and in a
collective bet for the whole civilization.  This would happen only
in the unlikely case when the size of civilization $N$ is not known
and at the same time the outcome of the measurement is correlated with
$N$.  When this situation does arise, the difference in predictions is
not due to any defect of the theory.  This difference is attributable
to the different information which was used as a basis for the
prediction.

\section{Comparing different theories}

\label{evidence}

The {\it evidence} for the theory $T$
inferred from the dataset $D$ is defined as the likelihood of the
data given the theory \cite{Jeffreys,Liddle},
\beq
E(T|D)=P(D|T).
\label{E}
\eeq
If there is a number of competing theories $T_a$, the quantity we
are interested in is the probability of a theory given the data,
\beq
P_a\equiv P(T_a|D)={\cal N}P(D|T_a)P(T_a).
\label{PNPP}
\eeq
Here, $P(D|T_a)=E_a$ is the evidence for $T_a$, $P(T_a)$ is the prior
probability for $T_a$, which reflects our prejudice for or against
this theory, and
\beq
{\cal N}=\left(\sum_a P(D|T_a)P(T_a)\right)^{-1}
\eeq
is a normalization constant.  The theory having the highest
probability $P_a$ should in principle be preferred.

The evidence $E$ is usually ranked on a logarithmic (Jeffreys) scale,
and the difference in $E$ for two competing theories is deemed
``significant'' if $1 < \Delta \ln E < 2.5$, ``strong'' if $2.5
<\Delta \ln E <5$ and ``decisive'' if $\Delta \ln E >5$
\cite{Jeffreys}.  One should keep in mind though that even if 
the evidence in favor of one of the theories is ``decisive'', this can
be counterbalanced by a strong enough bias against that theory in the
prior.

\subsection{Humans vs. Jovians}

Hartle and Srednicki \cite{HS} recently discussed an example intended to
show that the assumption of typicality in a wider reference class can
lead to absurd conclusions. This is an example where indexical information
plays an important role. Suppose we have a theory $T_1$ predicting
that there is no life on Jupiter and theory $T_2$ that says that
Jupiter is populated by intelligent beings and has population size $J$
much greater than the human population $H$ on Earth, $J\gg H$.
According to $T_1$, we are typical observers in the Solar system, and
according to $T_2$ we are not.  Hartle and Srednicki argue that it
would be absurd to dismiss $T_2$ merely because it makes us atypical.
We note, however, that the principle of mediocrity does not
necessarily imply that $T_2$ should be disfavored. 
This is because a smaller evidence can be compensated for by a higher
prior in Eq. (\ref{PNPP}).

The prior simply reflects our preference for the theory before we perform
the measurement, and to our knowledge there is no generally agreed upon
method for calculating it. For instance, one may adopt the so-called
self-indication assumption (SIA), which asserts that we should, all
other things being equal, assign a higher probability to the theory
predicting a larger number of observers.  SIA was first introduced by
Dieks \cite{Dieks} and further discussed by Bostrom \cite{Bostrom}.
Olum \cite{Olum} presented persuasive arguments that SIA should be
adopted, but it still remains somewhat controversial.\footnote{One
of the difficulties of the SIA is the so-called ``presumptuous
philosopher'' problem, as discussed in \cite{Bostrom}.}  The question
of validity of the SIA is peripheral to the issues we are concerned
with here.  For the sake of the argument, in this section we shall
adopt the SIA.

Suppose some individual (let us call him $K$) inspected himself enough
to tell that he is an intelligent observer in the Solar system, but he
does not know yet whether he is human or Jovian.  Then the principle
of mediocrity mandates that he regard himself as randomly picked from
the reference class $C_{HJ}$ including all human and Jovian observers
in the Solar system.  According to $T_1$ this class coincides with the
class $C_H$ of humans and has $H$ individuals, and according to $T_2$
it is much wider than $C_H$ and has $(J+H)$ individuals.  The
probability for $K$ to exist at all is proportional to the total number
of individuals.  Thus, having obtained the indexical information that
he exists, he should update the prior probabilities for the two
theories by enhancing the probability for $T_2$ by the factor
$(J+H)/H$ relative to that for $T_1$. 

Before taking into account 
any indexical information, we may characterize the two theories by prior probabilities 
$P(T_1)$ and $P(T_2)$. Then, according to the SIA, the fact of the observer's existence (which we
will also denote by $K$) updates such probabilities to $P(T_1|K)=P(T_1) H/(2H+J)$, 
and $P(T_2|K)=P(T_2)(H+J)/(2H+J)$.

Suppose now that $K$ looks out of the window and discovers that he is on
Earth. According to $T_2$ this second piece of indexical information has probability
$P(K\in H|T_2)=H/(J+H)$, and the updated probability for $T_2$ is
$$
P_2=P(T_2|K;K\in H)=P(T_2|K)P(K\in H|T_2)/P(K\in H),
$$
where $P(K \in H) = P(K \in H | T_1)P(T_1|K) +  P(K \in H | T_2) P(T_2|K)$
is the total probability for K to be human (after the first piece of
indexical information -that K exists- has been used).
On the other hand, the observation that $K$ is human is guaranteed in the theory $T_1$, $P(K\in H|T_1)=1$
and therefore 
$$P_1=P(T_1|K;K\in H)=P(T_1|K)/P(K\in H).$$
The net result is that $P_2/P_1=P(T_2)/P(T_1)$. In other words, the observation that $K$ is human
does not give him any information about the existence of life on Jupiter.

In conclusion, the principle of mediocrity does not necessarily make $T_2$ less likely than $T_1$
in spite of the fact that we are less typical in $T_2$ than in $T_1$. In the example above we have 
avoided this conclusion by using the self-indication assumption (SIA). We reiterate that this assumption 
is somewhat controversial, as is the more general problem of assigning prior probabilities to different 
theories. This is nevertheless an interesting subject which we leave for future discussion.

\section{Prediction vs. explanation}

Predicting the results of future measurements may be too narrow a goal
for a theory of the multiverse.  Many constants of Nature and
cosmological parameters have already been measured, so we have missed
the opportunity to predict them, but we can still try to explain their
values.

In order to explain the observed value of some parameter $\alpha$, we
can utilize the same approach as we outlined above, except we now have
to ignore the information we have about $\alpha$.  In other words, we
should consider a broader reference class $C^{(\alpha)}$, including
observers who measured the same values of the constants as we did,
except for the value of $\alpha$, which is allowed to vary from one
observer to another.  This reference class is then divided into
subclasses, $C^{(\alpha)}_j$, according to the measured values
$\alpha_j$, and the corresponding branching ratios are interpreted as
probabilities $P(\alpha_j)$.  The evidence for the underlying theory
can be evaluated using Eq.(\ref{E}), as before.  This evidence
can be used to discriminate between different models.  In
the same manner, one can introduce still broader reference classes,
$C^{(\alpha,\beta,...)}$, by allowing several parameters to vary.

As an illustration, we consider the case of the cosmological constant
$\Lambda$.  (For a review, see \cite{AV07}. More recent discussion can be 
found in \cite{TRWA,Levon2,Peacock}.)  In
order to explain the smallness of the observed value $\Lambda_*$, we
choose a reference class $C^{(\Lambda)}$ which includes all human-like
observers performing independent measurements of $\Lambda$ in parts of
the universe where all particle physics and cosmological parameters
other than $\Lambda$ have the same values as in our local region.
The corresponding distribution can be represented as 
\beq
P_1(\Lambda)\propto P_V(\Lambda)n_{obs}(\Lambda),
\label{PPf}
\eeq
where the volume distribution $P_V(\Lambda)$ characterizes the
fraction of (comoving) volume occupied by regions with a given value
of $\Lambda$ and $n_{obs}(\Lambda)$ is the number density of observers
belonging to the chosen reference class.

The standard argument suggests \cite{Efstathiou,AV96,Weinberg97} that
the volume distribution is well approximated by
\beq
P_V(\Lambda)\approx {\rm const},
\eeq
because the anthropic range where $n_{obs}(\Lambda)$ is appreciably
different from zero is much narrower than the full range of
$\Lambda$.  Note that a logarithmic distribution, which gives
equal likelihood to all orders of magnitude (the so-called Jeffreys
prior) would not be a suitable choice for $P_V(\Lambda)$. We know that
Lambda is a sum of very large positive and negative contributions
which are fine-tuned to near-cancellation.  Decreasing the order of
magnitude of $\Lambda$ corresponds to a higher level of fine-tuning
and should be given a lower probability.  The "standard argument" that
we referred to in the text assumes that $\Lambda = 0$ is not a special
point of the volume distribution.  (This point is singular for the
Jeffreys prior.)  Perhaps more importantly, the volume distribution
$P_V(\Lambda)$ is not a prior in the usual sense.  It is a claculable
distribution and has been demonstrated to be flat in some multiverse
models \cite{GV,Weinberg00,SPO,Shenker}.

For the anthropic factor $n_{obs}(\Lambda)$, the analysis of positive 
and negative $\Lambda$ is rather different; for our purposes it will 
be sufficient to focus on $\Lambda >0$.  Then the density of observers
is usually approximated as 
\beq
n_{obs}\propto F(M>M_G,\Lambda),
\label{nobs}
\eeq
where the quantity on the right-hand side is the asymptotic fraction
of baryonic matter which clusters into objects of mass greater than
the characteristic galactic mass $M_G$.  The idea here is that there
is a certain average number of stars per unit baryonic mass and
certain number of observers per star, and that these numbers are not
strongly affected by the value of $\Lambda$ in the range of interest.
Using the Press-Schechter formalism to approximate the fraction of
collapsed matter $F$ and combining Eqs.(\ref{PPf})-(\ref{nobs}), one
finds \cite{MSW,GLV}
\beq 
P_1(\Lambda)={\sqrt{\pi}\over{\Lambda_1}} {\rm
erfc}\left[\left(\Lambda /\Lambda_1\right)^{1/3}\right].
\label{P0}
\eeq
Here, 
\beq
\Lambda_1 \approx 2\rho_m\sigma_G^3,
\label{Lambda1}
\eeq
$\rho_m$ is the density of nonrelativistic matter and $\sigma_G$ is
the linearized density contrast on the galactic scale.  The prefactor
in (\ref{P0}) is chosen so that the distribution $P_1(\Lambda)$ is
properly normalized.  With the observed values of $\rho_m$ and
$\sigma_G$, the quantity $\Lambda_1$ in (\ref{Lambda1}) is comparable
to the observed value $\Lambda_*$, $\Lambda_1 \approx 3\Lambda_*$.

An alternative explanation for the observed value of $\Lambda$ is that
$\Lambda$ just is what it is.  It has to have some value and we have
found it.  There has been no anthropic selection.  In this approach,
$\Lambda$ has no reason to be smaller than the value set by the
supersymmetry breaking scale, $\Lambda_{SUSY}$.  No particular value
below that scale is preferred, so the prior distribution for $\Lambda$
can be set as $P_2(\Lambda)=\Lambda_{SUSY}^{-1}$ for
$\Lambda<\Lambda_{SUSY}$ and $P_2(\Lambda) =0$ for
$\Lambda>\Lambda_{SUSY}$.  Assuming that the measurement errors
are small, $\delta\Lambda\ll \Lambda_*$, we have
\beq
E_1/E_2=P_1(\Lambda_*)/P_2(\Lambda_*) = \sqrt{\pi}{\Lambda_{SUSY}
\over{\Lambda_1}}{\rm erfc}\left[\left(\Lambda_*/\Lambda_1 
\right)^{1/3}\right] \approx 0.2 {\Lambda_{SUSY}\over{\Lambda_*}}.
\eeq
We know that $\Lambda_{SUSY}\gtrsim (1 {\rm TeV})^4$, and thus
$E_1/E_2\gtrsim 10^{60}$.  Comparison of the two theories on the
Jeffreys scale gives $\Delta\ln E > 138$, and thus the evidence in
favor of the multiverse model is ``decisive''.

If one is not interested in explanation and wants to stick to the
business of prediction, the calculation should be somewhat different.
Imagine we are back in mid-1990's, when the value of $\Lambda$ was not
known, and we want to make a prediction for the upcoming supernova
measurements.  The current upper bound on $\Lambda$ at the time was
\cite{Fukugita}.  
\beq 
\Lambda\lesssim 4\rho_m.
\label{bound}
\eeq
The values of some relevant parameters, such as $\sigma_G$ were known
with a rather poor accuracy, but to simplify the discussion we shall
assume as before that all parameters other than $\Lambda$ have been
accurately measured.  

The quantity $4\rho_m$ is somewhat less than $\Lambda_1$ in
(\ref{Lambda1}); hence the distribution (\ref{P0}) in the range
(\ref{bound}) can be roughly approximated as flat, 
\beq
P(\Lambda) \approx {1\over{4\rho_m}} = {\rm const}.  
\eeq
The probability for $\Lambda$ to be smaller than any given value is
then $P(<\Lambda)\approx \Lambda/4\rho_m$.  For example, $P(<0.2
\rho_m) \approx 0.05$.  Thus, in this simplified model, one can
predict at 95\% confidence level that $\Lambda >0.2 \rho_m$.  This is
the essence of the prediction made in
Refs. \cite{Weinberg89,AV95,Efstathiou,MSW}.  We note that prediction
and explanation were not cleanly separated in
Refs.~\cite{Weinberg89,AV95,Efstathiou,MSW}.  Rather, they gave a
mixture of both.\footnote{Anthropic explanations for the smallness of
$\Lambda$ have been suggested earlier in \cite{Weinberg87,Linde87}.}

\section{Acknowledgements}

We would like to thank Alan Guth, Andrew Liddle, Ken Olum and Michael Salem for very
useful discussions.  We are also grateful to Misao Sasaki and the
Yukawa Institute for their hospitality at the workshop YITP-W-07-10,
where this work was originated.  This work was supported in part by
the Fundamental Questions Institute (JG and AV), by grants FPA2007-66665C02-02 and 
DURSI 2005-SGR-00082 (JG), and by the National Science Foundation (AV).

\end{document}